\documentclass[twocolumn,english,aps,amsmath,amssymb,superscriptaddress,showpacs,eprinv]{revtex4-2}
\usepackage[T1]{fontenc}
\usepackage[utf8]{inputenc}
\usepackage{color}
\usepackage{babel}
\usepackage{array}
\usepackage{bm}
\usepackage{multirow}
\usepackage{amsmath}
\usepackage{graphicx}
\usepackage[unicode=true,
 bookmarks=false,
 breaklinks=false,pdfborder={0 0 1},backref=false,colorlinks=false]
 {hyperref}
\hypersetup{
 colorlinks,linkcolor=blue,anchorcolor=blue,citecolor=blue,urlcolor=blue}

\makeatletter

\providecommand{\tabularnewline}{\\}

\usepackage{makecell}

\makeatother

\begin{document}
\title{Systematic investigation of emergent particles in type-III magnetic space groups}
\date{\today}
\author{Gui-Bin Liu}
\thanks{These two authors contribute equally to this work.}
\affiliation{Centre for Quantum Physics, Key Laboratory of Advanced Optoelectronic
Quantum Architecture and Measurement (MOE), School of Physics, Beijing
Institute of Technology, Beijing 100081, China\\ and Beijing Key
Laboratory of Nanophotonics and Ultrafine Optoelectronic Systems,
School of Physics, Beijing Institute of Technology, Beijing 100081,
China}
\author{Zeying Zhang}
\thanks{These two authors contribute equally to this work.}
\affiliation{College of Mathematics and Physics, Beijing University of Chemical
Technology, Beijing 100029, China}
\author{$\text{Zhi-Ming Yu}$}
\affiliation{Centre for Quantum Physics, Key Laboratory of Advanced Optoelectronic
Quantum Architecture and Measurement (MOE), School of Physics, Beijing
Institute of Technology, Beijing 100081, China\\ and Beijing Key
Laboratory of Nanophotonics and Ultrafine Optoelectronic Systems,
School of Physics, Beijing Institute of Technology, Beijing 100081,
China}
\author{Shengyuan A. Yang}
\affiliation{Research Laboratory for Quantum Materials, Singapore University of
Technology and Design, Singapore 487372, Singapore}
\author{Yugui Yao}
\email{ygyao@bit.edu.cn}

\affiliation{Centre for Quantum Physics, Key Laboratory of Advanced Optoelectronic
Quantum Architecture and Measurement (MOE), School of Physics, Beijing
Institute of Technology, Beijing 100081, China\\ and Beijing Key
Laboratory of Nanophotonics and Ultrafine Optoelectronic Systems,
School of Physics, Beijing Institute of Technology, Beijing 100081,
China}
\begin{abstract}
In three-dimensional (3D) crystals, emergent particles arise when
two or multiple bands contact and form degeneracy (band crossing)
in the Brillouin zone. Recently a complete classification of emergent
particles in 3D nonmagnetic crystals, which described by the type-II
magnetic space groups (MSGs), has been established. However, a systematic
investigation of emergent particles in magnetic crystals has not yet
been performed, due to the complexity of the symmetries of magnetically
ordered structures. Here, we address this challenging task by exploring
the possibilities of the emergent particles in the 674 type-III MSGs.
Based on effective $\bm{k}\cdot\bm{p}$ Hamiltonian and our classification
of emergent particles {[}\href{https://doi.org/10.1016/j.scib.2021.10.023}{Yu \textit{et al.}, Sci. Bull. \textbf{67}, 375 (2002)}{]},
we identify all possible emergent particles, including spinful and
spinless, essential and accidental particles in the type-III MSGs.
We find that all emergent particles in type-III MSGs also exist in
type-II MSGs, with only one exception, i.e. the combined quadratic
nodal line and nodal surface (QNL/NS). Moreover, tabulations of the
emergent particles in each of the 674 type-III MSGs, together with
the symmetry operations, the small corepresentations, the effective
$\bm{k}\cdot\bm{p}$ Hamiltonians, and the topological character of
these particles, are explicitly presented. Remarkably, combining this
work and our homemade \textsf{SpaceGroupIrep} and \textsf{MSGCorep}
packages will provide an efficient way to search topological magnetic
materials with novel quasiparticles.
\end{abstract}
\maketitle

\section{Introduction}

Since the discovery of topological Weyl and Dirac semimetals, the
investigation of emergent particles has experienced rapid development
and been attracting a variety of interests in condensed matter physics
\citep{Wan_Savrasov_2011_83_205101__Topological,young_dirac_2012,burkov_topological_2011,Wang_Fang_2012_85_195320__Dirac,Fang_Bernevig_2012_108_266802__Multi,Burkov_Balents_2011_107_127205__Weyl,Liu_Chen_2014_343_864__Discovery,young_dirac_2015,Xu_Hasan_2015_349_613__Discovery,soluyanov_type-ii_2015,Zhu_Soluyanov_2016_6_31003__Triple,Bradlyn_Bernevig_2016_353_5037__Dirac,chiu_classification_2016,Armitage_Vishwanath_2018_90_15001__Weyl,Li_Yang_2017_96_81106__Type,Yan_Wang_2017_96_41103__Nodal,Wu_Yang_2018_97_115125__Nodal,bi_nodal-knot_2017,Weng_Kawazoe_2015_92_45108__Topological,Ma_Yao_2018_98_201104__Mirror,Fu_Yao_2018_98_75146__Hourglasslike,hasan_weyl_2021,li_type-iii_2021}.
Compared with the elementary particles in high-energy physics, the
quasiparticles in solids have much more abundant species due to looser
symmetry constraints and then embrace more rich physics \citep{Wang_Fang_2013_88_125427__Three,lu_quantum_2017,Son_Spivak_2013_88_104412__Chiral,Gorbar_Shovkovy_2014_89_85126__Chiral,Bzdusek_Soluyanov_2016_538_75__Nodal,Ezawa_Ezawa_2016_116_127202__Loop,Bian_Hasan_2016_93_121113__Drumhead,Wieder_Kane_2016_116_186402__Double,yu_predicted_2016,yu_quadratic_2019,Bzdusek_Sigrist_2017_96_155105__Robust,Yu_Weng_2017_12_127202__Topological,chen_weak_2019,Nagaosa_Tokura_2020_5_621__Transport}.
Thus identifying and classifying all the possible emergent particles
in solids becomes a fundamentally important but also a challenging
work. Recently, in Ref. \citep{Yu_Yao_2021___2102.01517_Encyclopedia}
we present a complete list of emergent particles in three-dimensional
(3D) nonmagnetic crystals with ${\cal T}$-symmetry. In this work,
we establish such list for the 3D magnetic crystals belonging to type-III
magnetic space groups (MSGs).

In three dimensions, the crystal structure of materials are described
by the symmetry of space groups (SGs). By introducing magnetic order,
the crystals exhibit one more degree of freedom and then should be
described by MSGs. There are in total 1651 MSGs which are divided
into four types. The tpye-I MSGs are just the ordinary SGs, and do
not have any anti-unitary operation. In contrast, the general form
of the other three types of MSGs can be written 
\begin{equation}
M=S+\mathcal{A}S
\end{equation}
with $S$ a unitary subgroup with index 2 of the MSG $M$ and $\mathcal{A}$
an anti-unitary operation. $M$ can be constructed from an ordinary
SG $G$. When $S=G$, the MSGs can be further classified into two
types by whether ${\cal A}$ is time reversal symmetry ${\cal T}$
(type II) or a combined operation of ${\cal T}$ and a pure translation
(type IV). One then knows that the type-II MSGs have ${\cal T}$ symmetry
and are applied to nonmagnetic crystals. In type-III MSGs, $S$ is
a isotranslational (translationengleiche) subgroup of $G$ with index
$2$ and $\mathcal{A}$ is a combined operation containing ${\cal T}$
and an unitary (spatial) operation in $G-S$, making ${\cal A}S=\mathcal{T}(G-S)$.

It is clear that the symmetry of the type-III MSGs is lower than that
of the type-II MSGs and heretofore most studies on emergent particles
are in systems with ${\cal T}$ symmetry, i.e. the systems belonging
to the type-II MSGs. However, it should be noted that the emergent
particles also can appear in magnetic systems \citep{Wan_Savrasov_2011_83_205101__Topological,tang_dirac_2016,wang_large_2018,xu_topological_2018,liu_magnetic_2019,morali_fermi-arc_2019,nie_magnetic_2020}.
Actually, the original candidate for topological Weyl semimetal is
a magnetic material \citep{Wan_Savrasov_2011_83_205101__Topological}.
In Ref. \citep{tang_dirac_2016}, Tang \textit{et al}. predicted orthorhombic
antiferromagnet CuMnAs as a candidate of magnetic Dirac semimetal.
Moreover, novel emergent particles in magnetic materials with higher-order
dispersion or 1D manifold of degeneracy also has been unveiled in
previous works \citep{wang_antiferromagnetic_2017,wang_type-i_2018,Zhang_Yang_2021_103_115112__Magnetic}.
However, people still lack an overall and systematic understanding
about what types of emergent particles can exist in magnetic crystals
with various MSGs.

Towards this goal, in this work we perform an exhaustive investigation
of the emergent particles in type-III MSGs and compile an encyclopedia
for them. This is done for each of the 674 type-III MSGs, and the
lists of all possible emergent particles along with the symmetry conditions,
the effective $\bm{k}\cdot\bm{p}$ Hamiltonians, and the topological
characters of these particles for each type-III MSG is presented in
SM-SIII (Section SIII in the supplementary material \citep{SM}). The main results
are summarized in Tabs. \ref{tab:classification}, \ref{tab:complex}
and the Tables in SM-SI, corresponding to the list of all possible
emergent particles in type-III MSGs and a quantitative mapping between
the emergent particles and type-III MSGs, respectively. Our key findings
are the following. (i) According to our classification, there exist
total \textit{18} types of spinless emergent particles and \textit{19}
types of spinful emergent particles, as shown Tab. \ref{tab:classification}.
All these emergent particles can be realized in nonmagnetic systems.
(ii) The type-III MSGs can also host several kinds of complex emergent
particles, which are constituted by two different types of particles
(see Tab. \ref{tab:complex}). Remarkably, one of the complex quasiparticles,
namely QNL/NS {[}combined quadratic nodal line (QNL) and nodal surface
(NS){]}, only exists in magnetic systems. (iii) This encyclopedia
provides a platform for systematic research on emergent particles
by scanning all type-III MSGs. Compared with case-by-case study, the
systematic research can usually provide comprehensive knowledge, complete
inspection, and deep insights. Much important information can be and
only can be inferred from our work. For example, only with our classification,
one knows that in type-III MSGs the largest topological charge (Chern
number) for the nodal point is $|{\cal C}|=3$ and the largest order
of energy splitting for nodal line is quadratic. For comparison, the
former is $|{\cal C}|=4$ and the latter is cubic in type-II MSGs
\citep{Fang_Bernevig_2012_108_266802__Multi,cui_charge-four_2021}.
At last, for the complex particle QNL/NS, we also construct concrete
lattice model to demonstrate its existence and study its surface state.

Our work not only presents a complete classification and detailed
analysis of the emergent particles in type-III MSGs but also is useful
for searching novel topological magnetic materials with desired emergent
particles. For example, for a given magnetic crystals, when the first-principles
band structure are obtained, one can use our homemade \textsf{MSGCorep}
package \citep{MSGCorep} to calculate the small corepresentations
(coreps) of the degenerate bands. Then the species of the degeneracy
can be directly identified by looking up the tables in this encyclopedia.

\section{Rationale}

The approach to obtain the results in this work is similar to that
in Ref. \citep{Yu_Yao_2021___2102.01517_Encyclopedia}. We first calculate
the small coreps at all high-symmetry k-points and k-lines in the
Brillouin zone (BZ) of each of the 674 type-III MSGs based on our
homemade package \textsf{SpaceGroupIrep} \citep{Liu_Yao_2021_265_107993__SpaceGroupIrep}
and \textsf{MSGCorep} \citep{MSGCorep}. Both single-valued {[}for
spinless systems, without spin-orbit coupling (SOC){]} and double-valued
coreps (for spinful systems, with SOC) are considered. Consider a
type-III MSG $M$, which can be written as $M=S+\mathcal{T}(G-S)$
\citep{BCbook}. For a wave vector $\bm{k}$ in the BZ, its magnetic
little group (MLG), denoted by $M_{\bm{k}}$, is the subgroup of $M$
which is composed of the elements whose point parts leave $\bm{k}$
invariant, i.e. $M_{\bm{k}}=\{\,Q\,|\,Q\in M\,\&\,P(Q)\bm{k}\doteq\bm{k}\}$,
in which $P(Q)$ means the point part of $Q$ and $\doteq$ means
two wave vectors differ by a reciprocal lattice vector. Note that
$\text{\ensuremath{\mathcal{T}\bm{k}=-\bm{k}}},$ $P(Q)=R$ if $Q=\{R|\bm{t}\}$,
and $P(Q)=\mathcal{T}R$ if $Q=\mathcal{T}\{R|\bm{t}\}$. The MLG
$M_{\bm{k}}$ relates to the little group of $\bm{k}$ in $S$, denoted
by $S_{\bm{k}}$, in two ways: (i) $M_{\bm{k}}=S_{\bm{k}}$, in this
case there is no element $\{R|\bm{t}\}$ in $G-S$ which satisfies
$R\bm{k}\doteq-\bm{k}$ and hence $M_{\bm{k}}$ is unitary; (ii) $M_{\bm{k}}=S_{\bm{k}}+A$,
in this case $S_{\bm{k}}$ is the unitary subgroup of $M_{\bm{k}}$
and all elements in $A$ are anti-unitary with $|S_{\bm{k}}|=|A|$.
Then the small coreps of $M_{\bm{k}}$ can be calculated according
to the small representations of $S_{\bm{k}}$ \citep{BCbook,Liu_Yao_2021_265_107993__SpaceGroupIrep,MSGCorep}.
Here, we adopt the convention used in the book \citep{BCbook} to
describe MSG. The book uses the BNS notation \citep{Belov_Smirnova_1957_1_487__1651,Belov_Smirnova_1957_2_311__Shubnikov}
for MSG, but some MSGs are mistaken by the authors, which is also
mentioned in \citep{Tang_Wan_2021___2103.08477_Exhaustive}. We has
corrected the MSGs which are not compatible with the BNS definition
\citep{MSGCorep}.

With the coreps information, we identify all the possible degeneracies
including both essential and accidental degeneracies. For each degeneracy
(at a certain high-symmetry wave vector $\bm{k}_{0}$), we construct
the $\bm{k}\cdot\bm{p}$ Hamiltonians according to the symmetry constraints
\begin{equation}
\hspace{-0.5em}\begin{cases}
\mathcal{D}(Q)H(\bm{k})\mathcal{D}(Q)^{-1}=H(R\bm{k}), & \text{if }Q=\{R|\bm{t}\}\\
\mathcal{D}(Q)H^{*}(\bm{k})\mathcal{D}(Q)^{-1}=H(-R\bm{k}), & \text{if }Q=\mathcal{T}\{R|\bm{t}\}
\end{cases},
\end{equation}
where $\mathcal{D}(Q)$ is the unitary corep matrix of $Q$ for each
$Q\in M_{\bm{k}_{0}}$ and $\mathcal{D}(Q)$ can be either irreducible
(for essential degeneracy) or reducible (for accidental degeneracy).
Using the iteratively simplifying algorithm, $H(\bm{k})$ can be obtained
upto any specified order of $\bm{k}$ \citep{MagneticKP}. Here we
use the lowest order of $\bm{k}$ that is essential to make correct
classification of emergent particles. Most of the physical properties
of the degeneracies, such as energy dispersion and topological charge
can be directly inferred from the constructed effective Hamiltonian.
Finally, we classify all the band crossings by the standard of the
classification established in \citep{Yu_Yao_2021___2102.01517_Encyclopedia}
and the results are shown in Tabs. \ref{tab:classification} (refer
to the SM of \citep{Yu_Yao_2021___2102.01517_Encyclopedia} for the
details of each notation in Tab. \ref{tab:classification}), \ref{tab:complex},
and the tables in SM. It should be pointed out that Weyl points at
general k-points only need translation symmetries to protect them
and hence they are not involved in our classification, as stated in
\citep{Yu_Yao_2021___2102.01517_Encyclopedia}.

\begin{table*}
\caption{Classification and statistics of emergent particles in type-III MSGs.
Similar to ref. \citep{Yu_Yao_2021___2102.01517_Encyclopedia}, Abbr
is the abbreviation for the notation of emergent particle, $d_{m}$
is the dimension of the degeneracy manifold, $d$ is the degree of
degeneracy of the band crossing, Ld is the leading order of the band
splitting near the crossing, and $\mathcal{C}$ is the topological
charge (Chern number for nodal point or Berry phase for nodal line)
of the emergent particles. $N_{\text{ess}}$ ($N_{\text{acc}})$ is
the spinless particle's occurrence number in SM-SIIIA (SM-SIIIB) for
essential (accidental) degeneracy, and $N_{\text{ess}}^{\text{SOC}}$
($N_{\text{acc}}^{\text{SOC}})$ is similar but for spinful particles
in SM-SIIIC (SM-SIIID). The number in the parentheses is the number
of MSGs that host the particle.\label{tab:classification}}

\begin{ruledtabular}
\centering{}%
\begin{tabular}{lrcccccccc}
\multirow{2}{*}{Notation} & \multirow{2}{*}{Abbr} & \multirow{2}{*}{$d_{m}$} & \multirow{2}{*}{$d$} & \multirow{2}{*}{Ld} & \multirow{2}{*}{$|\mathcal{C}|$} & \multicolumn{4}{c}{Occurrence number}\tabularnewline
 &  &  &  &  &  & $N_{\text{ess}}$ & $N_{\text{acc}}$ & $N_{\text{ess}}^{\text{SOC}}$ & $N_{\text{acc}}^{\text{SOC}}$\tabularnewline
\hline 
Charge-1 Weyl point & C-1 WP & 0 & 2 & (111) & \Gape[6pt][0pt]{1} & 130 (59) & 1448 (321) & 218 (76) & 1448 (321)\tabularnewline
Charge-2 Weyl point & C-2 WP & 0 & 2 & (122) & 2 & 83 (37) & 228 (56) & 29 (21) & 228 (56)\tabularnewline
Charge-3 Weyl point & C-3 WP & 0 & 2 & (133) & 3 & $\times$ & 42 (14) & ${\color{red}\times}$ & 42 (14)\tabularnewline
Charge-4 Weyl point & C-4 WP & 0 & 2 & (223) & 4 & ${\color{red}\times}$ & $\times$ & $\times$ & $\times$\tabularnewline
 &  &  &  &  &  &  &  &  & \tabularnewline
Triple point & TP & 0 & 3 & (111) & -- & 47 (18) & 748 (222) & ${\color{red}\times}$ & 67 (27)\tabularnewline
\textcolor{black}{Charge-2 triple point} & C-2 TP & 0 & 3 & (111) & 2 & 15 (10) & $\times$ & 5 (5) & $\times$\tabularnewline
\textcolor{black}{Quadratic triple point} & QTP & 0 & 3 & (122) & -- & $\times$ & 28 (10) & $\times$ & $\times$\tabularnewline
\textcolor{black}{Quadratic contact triple point} & QCTP & 0 & 3 & (222) & 0 & 60 (26) & $\times$ & $\times$ & $\times$\tabularnewline
 &  &  &  &  &  &  &  &  & \tabularnewline
Dirac point & DP & 0 & 4 & (111) & 0 & 83 (59) & 173 (102) & 349 (161) & 565 (236)\tabularnewline
\textcolor{black}{Charge-2 Dirac point} & C-2 DP & 0 & 4 & (111) & 2 & 2 (2) & 30 (30) & 6 (4) & 30 (30)\tabularnewline
Charge-4 Dirac point & C-4 DP & 0 & 4 & (111) & 4 & $\times$ & $\times$ & ${\color{red}\times}$ & $\times$\tabularnewline
\textcolor{black}{Quadratic Dirac point} & QDP & 0 & 4 & (122) & 0 & 42 (30) & 18 (18) & 9 (7) & 10 (10)\tabularnewline
Charge-4 quadratic Dirac point & C-4 QDP & 0 & 4 & (122) & 4 & $\times$ & $\times$ & ${\color{red}\times}$ & $\times$\tabularnewline
\textcolor{black}{Quadratic contact Dirac point} & QCDP & 0 & 4 & (222) & 0 & $\times$ & $\times$ & 13 (10) & $\times$\tabularnewline
\textcolor{black}{Cubic Dirac point} & CDP & 0 & 4 & (133) & 0 & $\times$ & $\times$ & 1 (1) & $\times$\tabularnewline
\textcolor{black}{Cubic crossing Dirac point} & CCDP & 0 & 4 & (223) & 0 & 3 (3) & $\times$ & $\times$ & $\times$\tabularnewline
 &  &  &  &  &  &  &  &  & \tabularnewline
\textcolor{black}{Sextuple point} & SP & 0 & 6 & (111) & 0 & 8 (8) & $\times$ & 4 (4) & $\times$\tabularnewline
Charge-4 sextuple point & C-4 SP & 0 & 6 & (111) & 4 & $\times$ & $\times$ & ${\color{red}\times}$ & $\times$\tabularnewline
Quadratic contact sextuple point & QCSP & 0 & 6 & (222) & 0 & $\times$ & $\times$ & ${\color{red}\times}$ & $\times$\tabularnewline
 &  &  &  &  &  &  &  &  & \tabularnewline
\textcolor{black}{Octuple point} & OP & 0 & 8 & (111) & 0 & $\times$ & $\times$ & 3 (3) & $\times$\tabularnewline
 &  &  &  &  &  &  &  &  & \tabularnewline
Weyl nodal line & WNL & 1 & 2 & (11) & $\pi$ & 1510 (395) & 3525 (470) & 1243 (262) & 1082 (232)\tabularnewline
Weyl nodal line net & WNL net & 1 & 2 & (11) & $\pi$ & 1143 (274) & 1222 (294) & 685 (150) & 148 (57)\tabularnewline
Quadratic nodal line & QNL & 1 & 2 & (22) & 0 & 740 (173) & $\times$ & 61 (19) & $\times$\tabularnewline
Cubic nodal line & CNL & 1 & 2 & (33) & $\pi$ & $\times$ & $\times$ & ${\color{red}\times}$ & $\times$\tabularnewline
 &  &  &  &  &  &  &  &  & \tabularnewline
\textcolor{black}{Dirac nodal line} & DNL & 1 & 4 & (11) & 0 & 12 (4) & $\times$ & 223 (52) & 178 (51)\tabularnewline
\textcolor{black}{Dirac nodal line net} & DNL net & 1 & 4 & (11) & 0 & ${\color{red}\times}$ & $\times$ & 6 (2) & 8 (4)\tabularnewline
 &  &  &  &  &  &  &  &  & \tabularnewline
Nodal surface & NS & 2 & 2 & (1) & -- & 1257 (147) & $\times$ & 765 (94) & $\times$\tabularnewline
Nodal surface \textcolor{black}{net} & NS net & 2 & 2 & (1) & -- & 442 (62) & $\times$ & 168 (30) & $\times$\tabularnewline
\end{tabular}
\end{ruledtabular}

\end{table*}

In Tab.~\ref{tab:classification}, for each emergent particle, we
explicitly list its occurrence number at the high-symmetry momenta
of all type-III MSGs and also list the number of the type-III MSGs
hosting it. For both counting number, four cases are listed separately:
spinless essential particle, spinless accidental particle, spinful
essential particle, and spinful accidental particle. These data can
tell us which emergent particles are common and which ones are rare.
As can be seen: (i) C-1 WP (Charge-1 Weyl point) is very common and
WP becomes more and more rare as $|\mathcal{C}|$ increases. Statistics
further shows that C-2 WP only exists in tetragonal and hexagonal
systems, and C-3 WP only exists on the $\Delta$($00u$) line in hexagonal
systems. (ii) Accidental WP's are much more common than essential
WP's, which is also true for TP (Triple point), DP (Dirac point),
and C-2 DP. (iii) CDP (Cubic Dirac point), CCDP (\textcolor{black}{Cubic
crossing Dirac point}), and OP (\textcolor{black}{Octuple point})
are very rare, especially CDP only exists at the double-valued small
corep $(A)A_{7}A_{7}$ of MSG 192.251 ($P6/m'c'c'$) (see Appendix
\ref{sec:corep-label} for the corep label). (iv) WNL (Weyl nodal
line) is more common than C-1 WP, and WNL net is also very common.

Compared with the emergent particles in type-II MSGs \citep{Yu_Yao_2021___2102.01517_Encyclopedia},
6 types of emergent particles, i.e. C-4 WP, C-4 DP, C-4 QDP (Charge-4
quadratic Dirac point), C-4 SP (Charge-4 sextuple point), QCSP (Quadratic
contact sextuple point), and CNL (Cubic nodal line), do not exist
in type-III MSGs. Detailed differences are emphasized in red in Tabs.~\ref{tab:classification}
and \ref{tab:complex}, in which red cross means the particle exists
in type-II MSGs but not in type-III MSGs, red number means the particle
does not exist in type-II MSGs but exists in type-III MSGs, and black
number (cross) means the particle exists (do not exist) in both type-II
and type-III MSGs. Consequently, one finds that all non-complex particles
in type-III MSGs also exist in type-II MSGs, and the complex emergent
particle QNL/NS is the only one which exists in type-III MSGs but
not in type-II MSGs.

\begin{table*}
\caption{Complex emergent particles existing in type-III MSGs. The format of
this table is similar to Tab. \ref{tab:classification}.\label{tab:complex}}

\begin{ruledtabular}
\centering{}%
\begin{tabular}{lrccccc}
\multirow{2}{*}{Notation} & \multirow{2}{*}{Abbr} & \multirow{2}{*}{$d$} & \multicolumn{4}{c}{Occurrence number}\tabularnewline
 &  &  & $N_{\text{ess}}$ & $N_{\text{acc}}$ & $N_{\text{ess}}^{\text{SOC}}$ & $N_{\text{acc}}^{\text{SOC}}$\tabularnewline
\hline 
Combined WNL and NS & WNL/NS & \Gape[6pt][0pt]{2} & 68 (26) & $\times$ & 54 (22) & $\times$\tabularnewline
Combined WNL net and NS net & WNLs/NSs & 2 & 22 (9) & $\times$ & \textcolor{red}{12 (3)} & $\times$\tabularnewline
Combined QNL and NS & QNL/NS & 2 & \textcolor{red}{6 (3)} & $\times$ & \textcolor{red}{6 (3)} & $\times$\tabularnewline
Combined QNL and WNL net & QNL/WNLs & 2 & 166 (61) & $\times$ & ${\color{red}\times}$ & $\times$\tabularnewline
Combined QNL net and WNL net & QNLs/WNLs & 2 & 18 (18) & $\times$ & $\times$ & $\times$\tabularnewline
\end{tabular}
\end{ruledtabular}

\end{table*}

\begin{table*}
\caption{Part of the spinless emergent particles in MSG 56.370 excerpted from
the 56.370 tables in SM-SIIIA and SM-SIIIB. The line above the table
indicates the information about the notation of MSG (its unitary subgroup
in the parentheses), the Bravais lattice, the generators of the MSG,
whether $I\mathcal{T}$ exists, and whether SOC is considered. $\bm{k}$
is a high-symmetry k-point or k-line defined in the Tab. 3.6 of \citep{BCbook},
\textquotedblleft generators\textquotedblright{} are the point parts
for the generators of the MLG of $\bm{k}$, \textquotedblleft dim\textquotedblright{}
is the dimension of the corep, and \textquotedblleft matrices\textquotedblright{}
are the corep matrices of the MLG generators. The unitary matrices
$\lambda_{m},$ $\sigma_{p},$ and $\Gamma_{q}$ are defined in SM-SIV.
All $\bm{k}\cdot\bm{p}$ Hamiltonians are defined in SM-SV. Node type
is just the type of emergent particles.\label{tab:56.370part}\\[-5pt]}

\textbf{56.370}, $Pc'cn'$ (14, $P2_{1}/c$)\hfill{}$\Gamma_{o},$
$\{C_{2y}|\frac{1}{2}\frac{1}{2}0\},$ $\{I|\frac{1}{2}\frac{1}{2}\frac{1}{2}\},$
$\{C_{2x}\mathcal{T}|\frac{1}{2}\frac{1}{2}0\}$, Without $I\mathcal{T}$,
without SOC\\

\begin{ruledtabular}
\centering{}%
\begin{tabular}{ccclllllc}
\multicolumn{2}{c}{$\bm{k}$} & \multirow{2}{*}{generators} & \multicolumn{3}{c}{corep} & \hspace{1.8em}$\bm{k}\cdot\bm{p}$  & node  & \multirow{2}{*}{$|\mathcal{C}|$}\tabularnewline
\cline{1-2} \cline{2-2} \cline{4-6} \cline{5-6} \cline{6-6} 
name  & info  &  & label  & dim  & matrices  & Hamiltonian  & type  & \tabularnewline
\hline 
$\Gamma$  & 000  & $C_{2y},I,C_{2x}\mathcal{T}$  & $(\Gamma)\Gamma_{1}^{+}$  & 1  & $1,1,1$  & $H_{10.46}^{(\Gamma)\Gamma_{1}^{+}}$  & --  & \tabularnewline
 &  &  & $(\Gamma)\Gamma_{1}^{-}$  & 1  & $1,-1,-1$  & $H_{10.46}^{(\Gamma)\Gamma_{1}^{+}}$  & --  & \tabularnewline
 &  &  & $(\Gamma)\Gamma_{2}^{+}$  & 1  & $-1,1,-1$  & $H_{10.46}^{(\Gamma)\Gamma_{1}^{+}}$  & --  & \tabularnewline
 &  &  & $(\Gamma)\Gamma_{2}^{-}$  & 1  & $-1,-1,1$  & $H_{10.46}^{(\Gamma)\Gamma_{1}^{+}}$  & --  & \tabularnewline
$Y$  & $\bar{\frac{1}{2}}00$  & $C_{2y},I,C_{2x}\mathcal{T}$  & $(Y)Z_{1}$  & 2  & $\sigma_{4},\sigma_{3},\sigma_{3}$ & $H_{50.282}^{(X)A_{2}}$  & P-WNL  & \tabularnewline
$U$  & $0\frac{1}{2}\frac{1}{2}$  & $C_{2y},I,C_{2x}\mathcal{T}$  & $(U)A_{1}A_{1}$  & 4  & $\Gamma_{49},-\Gamma_{70},-\Gamma_{10}$  & $H_{56.370}^{(U)A_{1}A_{1}}$  & DP  & 0\tabularnewline
$S$  & $\bar{\frac{1}{2}}\frac{1}{2}0$  & $C_{2y},I,C_{2x}\mathcal{T}$  & $(S)C_{1}$  & 2  & $\sigma_{4},\sigma_{3},\sigma_{4}$  & $H_{55.358}^{(X)B_{2}}$  & P-NS  & \tabularnewline
$R$  & $\bar{\frac{1}{2}}\frac{1}{2}\frac{1}{2}$  & $C_{2y},I,C_{2x}\mathcal{T}$  & $(R)E_{1}^{+}E_{2}^{+}$  & 2  & $i\sigma_{3},-\sigma_{0},-i\sigma_{4}$  & $H_{55.358}^{(S)D_{1}^{+}D_{2}^{+}}$  & P-WNL/NS  & \tabularnewline
 &  &  & $(R)E_{1}^{-}E_{2}^{-}$  & 2  & $i\sigma_{3},\sigma_{0},i\sigma_{4}$  & $H_{55.358}^{(S)D_{1}^{+}D_{2}^{+}}$  & P-WNL/NS  & \tabularnewline
$D$  & $XS$  & $C_{2y},C_{2x}\mathcal{T}$  & $(D)W_{1}W_{2}$  & 2  & $\lambda_{20}\sigma_{3},\lambda_{20}\sigma_{4}$  & $H_{55.358}^{(D)V_{1}V_{2}}$  & L-NS  & \tabularnewline
$E$  & $TR$  & $\sigma_{y},C_{2z}\mathcal{T}$  & $(E){\rm UN_{1}UN_{2}}$  & 2  & $-i\sigma_{3},\sigma_{1}$  & $H_{27.80}^{(Z)B_{1}B_{2}}$  & WNL  & $\pi$\tabularnewline
\multicolumn{9}{c}{\Gape[6pt]{Accidental degeneracy on high-symmetry k-line}}\tabularnewline
$\Delta$  & $\Gamma Y$  & $C_{2y},C_{2x}\mathcal{T}$  & $\{(\Delta)\Lambda_{1},(\Delta)\Lambda_{2}\}$  & 2  & $\lambda_{20}\sigma_{3},\lambda_{20}\sigma_{3}$  & $H_{49.270}^{\{(\Delta)\Lambda_{1},(\Delta)\Lambda_{2}\}}$  & C-1 WP  & 1\tabularnewline
$\Sigma$  & $\Gamma X$  & $\sigma_{y},C_{2z}\mathcal{T}$  & $\{(\Sigma){\rm UN}_{1},(\Sigma){\rm UN}_{2}\}$  & 2  & $\sigma_{3},\sigma_{0}$  & $H_{25.59}^{\{(\Sigma){\rm UN}_{1},(\Sigma){\rm UN}_{2}\}}$  & P-WNL  & \tabularnewline
\end{tabular}
\end{ruledtabular}

\end{table*}

\section{An example: MSG 56.370}

As discussed above, we explore all the possibilities of the emergent
particles in type-III MSGs and tabulate the results one MSG by MSG.
The resulted tables are listed in SM. We then use a spinless system
with MSG No. 56.370 ($Pc'cn')$ as an example to provide a glimpse
of the encyclopedia. Tab.~\ref{tab:56.370part} is an example excerpted
from the tables for MSG 56.370 in SM-SIIIA and SM-SIIIB. In Tab.~\ref{tab:56.370part},
the first line provides some basic information of MSG 56.370, including
unitary subgroup, BZ type, generating elements, whether the MSG has
$I\mathcal{T}$ symmetry (combined spatial inversion symmetry $I$
and $\mathcal{T}$), and whether SOC effect is considered.

The main part of Tab.~\ref{tab:56.370part} can be divided into six
parts: information of high-symmetry momentum $\bm{k}$, the point
parts of the generating elements of $M_{\bm{k}}$, the corep information
of $M_{\bm{k}}$, and the effective Hamiltonian, the type and the
topological charge of the degeneracies. Particularly, we find that
many coreps share the same matrices and so do the $\bm{k}\cdot\bm{p}$
Hamiltonians. Thus, we also explicitly list all the possible corep
matrices and Hamiltonians in SM-SIV and SM-SV respectively. For example,
the corep matrices for the generators of $M_{\bm{k}=Y}$ are $\mathcal{D}($$\{C_{2y}|\frac{1}{2}\frac{1}{2}0\})=\sigma_{4}$
and $\mathcal{D}($$\{I|\frac{1}{2}\frac{1}{2}\frac{1}{2}\})=\mathcal{D}(\mathcal{T}\{C_{2x}|\frac{1}{2}\frac{1}{2}0\})=\sigma_{3}$
\citep{getMLG} with
\[
\sigma_{4}=\begin{pmatrix}0 & 1\\
-1 & 0
\end{pmatrix},\ \ \sigma_{3}=\begin{pmatrix}1 & 0\\
0 & -1
\end{pmatrix},
\]
and the effective Hamiltonian is 
\[
H_{50.282}^{(X)A_{2}}=c_{2}k_{x}\sigma_{1}-c_{3}k_{y}\sigma_{2}+c_{1}\sigma_{0}.
\]
The subscript and superscript in $H_{50.282}^{(X)A_{2}}$ means that
this matrix form of effective Hamiltonian firstly appear for the degeneracy
with corep $(X)A_{2}$ at MSG 50.282.

\begin{figure*}
\begin{centering}
\includegraphics[width=16cm]{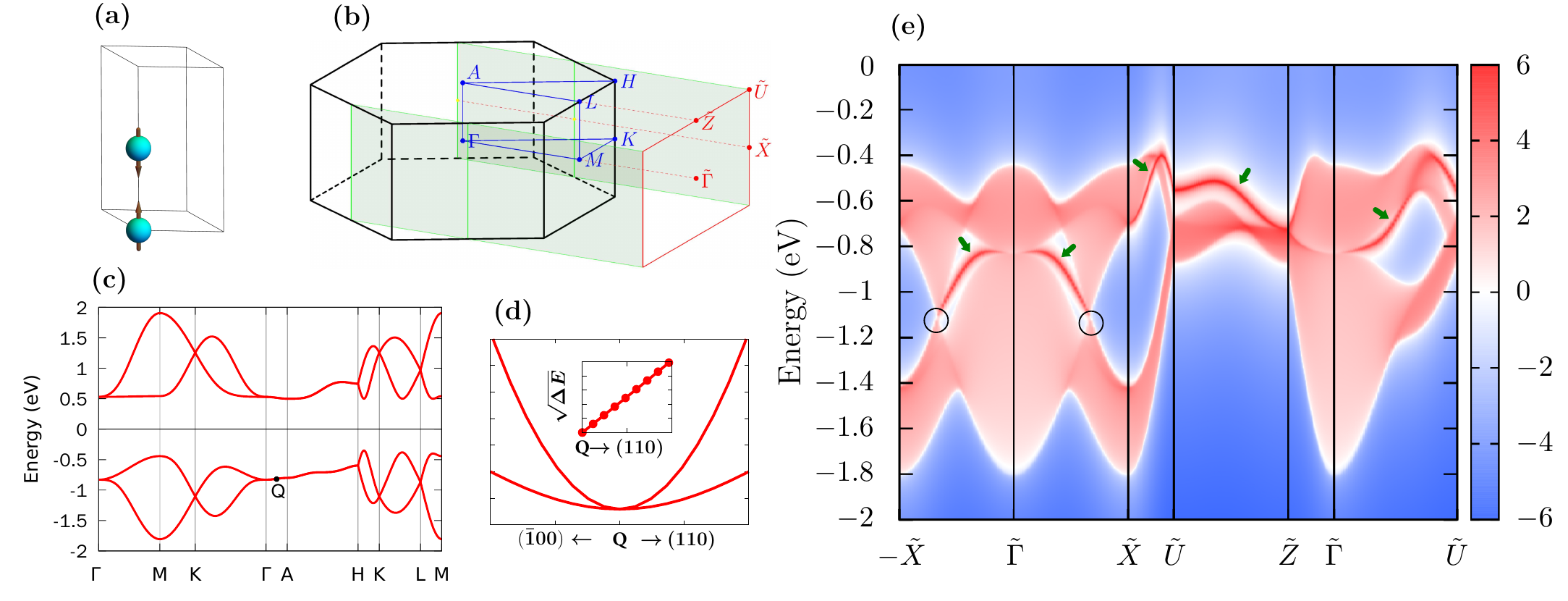}
\par\end{centering}
\caption{Unit cell (a), bulk BZ and its projection to (100) surface (b), bulk
energy bands (c,d), and density of states of semi-infinite system
with (100) surface (e) of the TB model of MSG 176.147. $\Delta E$
in (d) means the band splitting. The parameters used are $\varepsilon=-0.5$,
$t_{1}=0.4$, $t_{2}=-0.1$, $t_{3}=0$, $r=-0.8$, $s_{1}=0.2$,
$s_{2}=s_{3}=s_{4}=0$. \label{fig:fig1}}
\end{figure*}

P-WNL (P-NS) indicates that the high-symmetry k-point (e.g. $Y$,
$S$) or the accidental band-crossing point on a high-symmetry k-line
(e.g $\Sigma$) is actually a point residing on a WNL (NS), and similarly
L-NS indicates that the high-symmetry k-line (e.g. $D$) resides on
a NS. Therefore, when counting the occurrence numbers in Tabs.~\ref{tab:classification}
and \ref{tab:complex}, P-WNL is regarded as WNL, P-WNL/NS is regarded
as WNL/NS, and both P-NS and L-NS are regarded as NS.

\section{Complex particle: QNL/NS}

As mentioned above, QNL/NS is a novel complex emergent particle that
does not exist in nonmagnetic systems. Here we take a spinful system
hosting QNL/NS for example to demonstrate the existence of QNL/NS.
There are three type-III MSGs hosting QNL/NS, i.e. 176.147 ($P6_{3}'/m'$),
193.259 ($P6_{3}'/m'cm'$), and 194.268 ($P6_{3}'/m'm'c$). Without
loss of generality, we construct a tight-binding (TB) model under
the symmetric constraints of MSG 176.147 by our homemade \textsf{MagneticTB}
package \citep{Zhang_Yao_2021___2105.09504_MagneticTB}. The simplest
$s$ orbitals, i.e. $|s\uparrow\rangle$ and $|s\downarrow\rangle$,
are adopted to construct the TB model, which is enough to capture
QNL/NS. Fig. \ref{fig:fig1}(a) is the unit cell with irrelevant atoms
omitted, showing an A-type antiferromagnetic configuration. The TB
Hamiltonian is given in Appendix \ref{sec:TB}.

The energy bands of the model is shown in Fig. \ref{fig:fig1}(c),
from which we can see that line $\Gamma$-$A$, line $A$-$H$ (in
fact the whole plane $AHL$) and point $K$ are doubly degenerate.
The four bands are divided into upper and lower portions which are
separated by a gap and possess the same degeneracy over the whole
BZ. Accordingly, the lower two bands are enough to host all possible
essential particles in MSG 176.147, including three non-complex types,
i.e. a QNL along $\Gamma$-$A$, a NS on plane $AHL$, and two C-1
WP's at $\pm K$ points, and a complex type QNL/NS at $A$ point (cf.
the 176.147 table in the SM-SIIIC). The band splitting around any
point on $\Gamma$-$A$, such as the $Q$ point in Fig. \ref{fig:fig1}(c),
is quadratic along both ($\bar{1}$00) and (110) directions, as shown
in Fig. \ref{fig:fig1}(d), which indicates $\Gamma$-$A$ is a QNL.
Fig. \ref{fig:fig1}(e) shows the density of states of a semi-infinite
system terminated by the (100) plane, in which the two circles indicate
the projection of the two C-1 WP's at $\pm K$ onto the (100) surface
BZ {[}see Fig. \ref{fig:fig1}(b){]} and the arrows indicate the surface
states.

It should be pointed out that the coexistence of both QNL along $\Gamma$-$A$
and NS on plane $AHL$ does not necessarily lead to QNL/NS at $A$
point. The existence of QNL/NS requires that the coreps of QNL and
NS satisfy the compatibility relation with the coreps of their intersection
point. For example, the type-II MSG 176.144 $(P6_{3}/m1'$) have both
QNL along $\Gamma$-$A$ and NS on plane $AHL$ but does not have
QNL/NS at $A$ point \citep{yu_quadratic_2019,Wu_Yang_2018_97_115125__Nodal}.
None of type-II MSGs satisfies such compatibility relation and this
is why QNL/NS does not exist in type-II MSGs.

\section{Conclusions}

In conclusion, we have studied all the possible emergent particles
that can be stabilized by type-III MSG symmetries and complied the
results to an encyclopedia. Band crossings at all high-symmetry k-points
and k-lines and originated from both single-valued and double-valued
small coreps are analyzed. In addition to the essential degeneracy
protected by a single small corep, accidental degeneracy induced by
a pair of small coreps is also considered. Compared with the results
of type-II MSGs in \citep{Yu_Yao_2021___2102.01517_Encyclopedia},
the non-complex emergent particles existing in type-III MSGs form
a subset of those in type-II MSGs, missing C-4 WP, C-4 DP, C-4 QDP,
C-4 SP, QCSP, and CNL in type-III MSGs. The complex particle QNL/NS
is the only one which exists in type-III MSGs but not in type-II MSGs.
One can easily check which type-III MSGs can host a certain emergent
particle in SM-SI. Apart from the emergent particles, this encyclopedia
provide a quick reference of all small coreps at/on every high-symmetry
k-point/k-line defined in \citep{BCbook}, and it also provides the
$\bm{k}\cdot\bm{p}$ Hamiltonians for these k-points/k-lines. This
work will be a convenient reference for emergent particles, symmetries,
and $\bm{k}\cdot\bm{p}$ models in the studies of magnetic topological
nodal materials and related fields, and it will also facilitate the
search for required emergent particles in magnetic materials.

\textcolor{red}{\textbf{Note: Please find the supplementary material ``MSGIII\_emergent\_SM.pdf'' in the source file (gzipped tar file).}}

\begin{acknowledgments}
YY acknowledges the support by the National Key R\&D Program of China
(Grant No. 2020YFA0308800), the NSF of China (Grants Nos. 11734003,
12061131002), and the Strategic Priority Research Program of Chinese
Academy of Sciences (Grant No. XDB30000000). GBL acknowledges the
support by the international cooperation project of NSF of China (Grant
No. 52161135108),the National Key R\&D Program of China (Grant No.
2017YFB0701600) and the Beijing Natural Science Foundation (Grant
No. Z190006). ZZ acknowledges the support by the NSF of China (Grant
No. 12004028), and the China Postdoctoral Science Foundation (Grant
No. 2020M670106). ZMY acknowledges the support by the NSF of China
(Grant No. 12004035).
\end{acknowledgments}

\appendix

\section{label of small corep \label{sec:corep-label}}

The small corep of a type-III MSG $M=S+\mathcal{T}(G-S)$ is constructed
from \textit{one} or \textit{two} of the small representations of
its unitary subgroup $S$. The BZ and k-point naming of $M$ are the
same with those of $G$, but they may be different from those of $S$.
Taking MSG 118.309 ($P\bar{4}'n'2$) for example, its $S$ subgroup
is space group No. 21 ($C222)$. Hence $M$ has a simple tetragonal
BZ, while $S$ has a base-centered orthogonal BZ. The k-point $A$
for $M$ corresponds to the k-point $T$ for $S$. Accordingly, we
use the label $(A)T_{1}T_{3}$ to describe the small corep of the
MLG $M_{\bm{k}(=A)}$ which is constructed from two small representations
$T_{1}$ and $T_{3}$ of the little group $S_{\bm{k}(=T)}$. For the
labels of small representations, such as the $T_{1}$ and $T_{3}$
here, the BC convention is adopted \citep{BCbook,Liu_Yao_2021_265_107993__SpaceGroupIrep}.
The symbol $A$ in the parentheses explicitly indicates the k-point
name for $M$, which is different from the k-point name $T$ for $S$
here. However, even if the k-point names for both $M$ and $S$ happen
to be the same, we still keep the parentheses. For example, MSG 118.309
also has small coreps $(A)T_{2}$, $(V)H_{1}H_{1}$, $(\Lambda)\Lambda_{1}$,
$(Z)Z_{2}Z_{2}$, $(R)R_{1}R_{2}$, and so on.

\section{TB model of MSG 176.147\label{sec:TB}}

According to the symmetries in MSG 176.147, a TB model based on orbitals
$\{|s\uparrow\rangle,|s\downarrow\rangle$\} at each of the two magnetic
atoms in the hexagonal cell shown in Fig. \ref{fig:fig1}(a) can be
constructed by the \textsf{MagneticTB} package \citep{Zhang_Yao_2021___2105.09504_MagneticTB}
to show the existence of QNL/NS emergent particle. The obtained Hamiltonian
is

\begin{equation}
H(\bm{k})=\begin{bmatrix}\varepsilon\sigma_{3}+h_{1} & h_{2}+h_{3}\\
h_{2}^{+}+h_{3}^{+} & \varepsilon\sigma_{3}+h_{1}^{*}
\end{bmatrix}
\end{equation}

\begin{equation}
h_{1}=\begin{bmatrix}t_{2}f_{1}(\bm{k}_{\parallel}) & t_{1}f_{2}(\bm{k}_{\parallel})\\
t_{1}^{*}f_{2}^{*}(\bm{k}_{\parallel}) & t_{3}f_{1}(\bm{k}_{\parallel})
\end{bmatrix}
\end{equation}

\begin{equation}
h_{2}=r\cos\frac{k_{z}}{2}\sigma_{1}
\end{equation}
\begin{equation}
h_{3}=\begin{bmatrix}s_{3}f_{2}(\bm{k}_{\parallel})\cos\frac{k_{z}}{2} & s_{1}f_{3}(\bm{k})+s_{2}f_{4}(\bm{k})\\
s_{1}f_{4}(\bm{k})+s_{2}f_{3}(\bm{k}) & s_{4}f_{2}^{*}(\bm{k}_{\parallel})\cos\frac{k_{z}}{2}
\end{bmatrix}
\end{equation}
in which $h_{1},h_{2},$ and $h_{3}$ are blocks from the 1st-, 2nd-,
and 3rd-neithbour hoppings respectively, $\varepsilon$ is a real
parameters, and $t_{i}$, $r,$ and $s_{j}$ are complex hopping parameters
(except $t_{2}$ and $t_{3}$ which are real). $\bm{k}_{\parallel}=(k_{x},k_{y})$,
and $f_{i}$ is defined as follows 
\begin{equation}
f_{1}(\bm{k}_{\parallel})=\cos k_{x}+\cos k_{y}+\cos(k_{x}+k_{y})
\end{equation}
\begin{equation}
f_{2}(\bm{k}_{\parallel})=\cos k_{x}+e^{-i\frac{2\pi}{3}}\cos k_{y}+e^{-i\frac{4\pi}{3}}\cos(k_{x}+k_{y})
\end{equation}
\begin{align}
f_{3}(\bm{k})= & \cos(k_{x}-\frac{k_{z}}{2})+\cos(k_{y}-\frac{k_{z}}{2})\nonumber \\
 & +\cos(k_{x}+k_{y}+\frac{k_{z}}{2})
\end{align}
\begin{align}
f_{4}(\bm{k})= & \cos(k_{x}+\frac{k_{z}}{2})+\cos(k_{y}+\frac{k_{z}}{2})\nonumber \\
 & +\cos(k_{x}+k_{y}-\frac{k_{z}}{2})
\end{align}
Compared with the original output of \textsf{MagneticTB}, in order
to make the result tidy, we have adjusted the bases by the transformation
matrix $\sigma_{0}\oplus\sigma_{1}$ and substituted for the original
parameters, i.e. $\varepsilon=(\text{e2}-\text{e1})/2$, $t_{1}=2[\text{t1}+\frac{i}{\sqrt{3}}(\text{t1}+2\text{t3})]$,
$t_{2}=2\text{t4}$, $t_{3}=2\text{t2}$, $r=2(\text{r2}-i\,\text{r1}$),
$s_{1}=2(\text{s5}+i\,\text{s1})$, $s_{2}=2(\text{s8}+i\,\text{s2})$,
$s_{3}=4[\text{s6}+\frac{i}{\sqrt{3}}(2\text{s3}+\text{s6})]$, and
$s_{4}=4[\text{s7}-\frac{i}{\sqrt{3}}(2\text{s4}+\text{s7})]$.


\begin{thebibliography}{58}
\bibitem [{get()}]{getMLG}%
  \BibitemOpen
  \href@noop {} {\bibinfo {title} {{The commands
  \texttt{getMLGElem[\{56,370\},"Y"]} and \texttt{showMLGCorep[\{56,370\},"Y"]}
  in the package \textsf{MSGCorep} \cite{MSGCorep} can easily give repsectively
  the elements and coreps of this MLG.}}}\BibitemShut {Stop}%
\end{thebibliography}
%

\end{document}